\documentclass[sigconf,nonacm,natbib=true]{acmart}
\AtBeginDocument{%
  }

\usepackage[T1]{fontenc}
\usepackage{graphicx}
\usepackage{subcaption}
\usepackage{xcolor}
\usepackage{xurl} 

\newcommand{\systemname}{\textsc{IIRSim Studio}}

\begin{document}

\title[\systemname{}: A Dashboard for User Simulation]{\systemname{}: A Dashboard for User Simulation}

\author{Saber Zerhoudi}
\orcid{0000-0003-2259-0462}
\affiliation{%
  \institution{University of Passau}
  \city{Passau}
  \country{Germany}
}
\email{saber.zerhoudi@uni-passau.de}

\author{Adam Roegiest}
\orcid{0000-0003-1265-8881}
\affiliation{%
  \institution{Zuva}
  \city{Toronto}
  \country{Canada}
}
\email{adam@roegiest.com}

\author{Michael Granitzer}
\orcid{0000-0003-3566-5507}
\affiliation{%
  \institution{University of Passau}
  \city{Passau}
  \country{Germany}
}
\email{michael.granitzer@uni-passau.de}

\makeatletter
\setlength{\skip\footins}{9pt plus 2pt minus 1pt}

\newcommand{\acmrightssize}{\fontsize{8}{9.5}\selectfont}

\setlength{\emergencystretch}{1.5em} 

\settopmatter{printacmref=false}

\newcommand{\firstpagerights}[1]{%
  \begingroup
    \renewcommand\thefootnote{}%
    \footnotetext{%
      \acmrightssize
      \raggedright
      \setlength{\parskip}{0pt}%
      \setlength{\parindent}{0pt}%
      #1%
    }%
    \addtocounter{footnote}{0}%
  \endgroup
}
\makeatother


\begin{abstract}
User simulation is a valuable methodology for evaluation in Information Retrieval (IR), enabling low-cost experimentation and counterfactual analysis. However, existing simulation frameworks are primarily code-centric libraries that require substantial setup effort, which limits adoption and hinders reproducibility. The bottleneck is not the simulation engines themselves, but the lack of infrastructure connecting experiment design, execution, and sharing into a single verifiable workflow. This paper introduces \systemname{},~\footnote{\label{fn:searchsim}\url{https://iirsim.com}} a web-based workbench that addresses this gap through four contributions: (1) a visual environment for composing simulation pipelines on top of simulation frameworks, serving both novices learning simulation concepts and experts piloting large-scale experiments; (2) a component lifecycle that supports authoring, versioning, and sharing custom simulation components through Git-backed storage and runtime injection; (3) a provenance model based on experiment bundles and environment templates that makes the scope of replication explicit; and (4) a shared-task workflow, demonstrated through the re-deployment of a Sim4IA micro-task. \systemname{} is available as a hosted service and as a portable containerized deployment.
\end{abstract}



\maketitle
\enlargethispage{2\baselineskip}
\firstpagerights{%
  © ACM, 2026. This is the author's version of the work.\\
  The definitive version was published in:
  \emph{Proceedings of the 49th International ACM SIGIR Conference on Research and Development in Information Retrieval (SIGIR '26), July 20--24, 2026, Melbourne, VIC, Australia}.\\
  DOI: \url{https://doi.org/10.1145/3805712.3808593}
}

\section{Introduction}
Interactive Information Retrieval (IIR) evaluation aims to reflect how users issue queries, examine results, click, reformulate, and stop. Cranfield-style test collections support repeatable experiments, but they often do not model the feedback loop between users and the system. Human studies capture interaction, yet they require time, resources, and can be hard to replicate. User simulation offers a controlled alternative that models sequences of user actions and supports repeated runs under fixed settings~\cite{Balog:2025:ArXiv,BalogZ:2025:ArXiv,Zhang:2017:ICTIR-Formal-Framework}.

Simulation frameworks in IR have matured~\cite{engelmann2024contextdriveninteractivequerysimulations,erbacher2022stateartusersimulation}, including modular engines such as SimIIR~\cite{Maxwell:2016:SIGIR-SimIIR,Zerhoudi:2022:CIKM-SimIIR2,Azzopardi:2024:SIGIR-AP} and recent work on conversational search simulation~\cite{Kiesel:2024:ECIR,Bernard:2025:UserSimCRS,Afzali_2023,Bernard2024::WSDM,deWit2023::Conversations}. Despite this progress, many simulation studies still rely on local scripts, hand-built environments, and one-off configuration files. The bottleneck is not the engines themselves but the absence of infrastructure that connects experiment design, execution, and sharing into a single verifiable workflow. New users face a steep learning curve that slows adoption, and replication requires reconstructing the environment, the component versions, and other settings using incomplete records.\footnote{A task for which Cranfield-style evaluation has also struggled with despite numerous attempts~\cite{Clancy:2019:OSIRRC,Frobe:2023:TireX,Armstrong:2009:EvaluatIR}.}

This paper introduces \systemname{}, a resource that reduces many of the associated costs for simulation experiments using the SimIIR engine~\cite{Azzopardi:2024:SIGIR-AP}. \systemname{} is not a new simulator model. It is a workbench and orchestration layer that supports composing a simulator pipeline, running it in a controlled container environment, inspecting logs, and exporting a versioned experiment bundle that others can rerun or scale up for more exhaustive experiments.

\begin{figure*}[t!]
\begin{center}
\includegraphics[width=\textwidth]{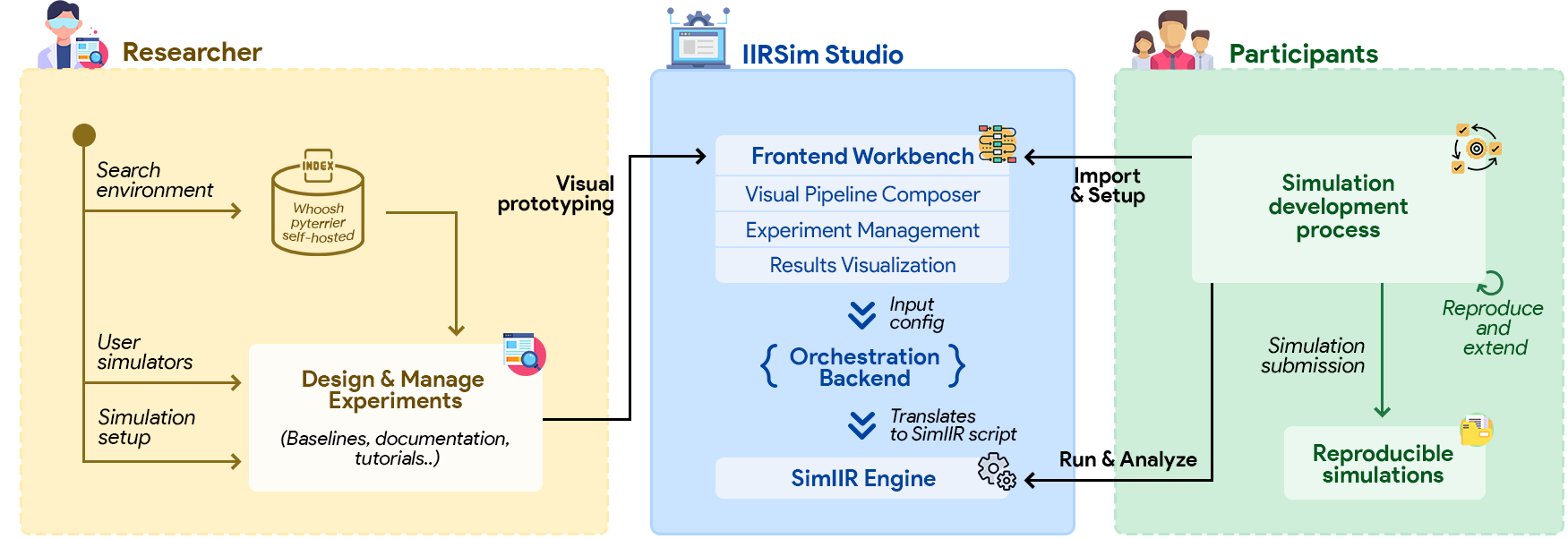}
\vspace{-3mm}
  \caption{The \systemname{} architecture. The Frontend Workbench provides visual pipeline composition, tutorials, and a playground for prototyping. The Orchestration Backend translates pipelines into versioned experiment bundles that are executed in isolated Docker containers, either through the workbench or at scale through the API wrapper.}
  \label{fig:architecture}
\end{center}
\end{figure*}

\systemname{} is designed for researchers with little simulation experience that may also be less technically adept, while also aiding experienced practitioners in developing and piloting more extensive experiments. The core design goals are: (1) interactive use through a visual interface for composing simulations (Section~\ref{sec:frentend-workbench}); (2) extensibility through custom components that can be authored and versioned (Section~\ref{sec:custom_components}); and (3) portability across hosted and local deployments (Section~\ref{sec:architecture}). Beginners learn core concepts through a hands-on tutorial and run short experiments in a \textit{Playground}. More advanced users can add custom components (e.g., novel Generative AI-based user models), manage experiment variants, and run large batches through an API wrapper or local deployment. With support for a wide spectrum of practitioner experience levels, \systemname{} also helps to facilitate shared tasks and their organization through a task environment that can be pinned and shared, and with participant submissions represented as rerunnable simulator configurations~\cite{Ferro:2018:JDIQ-Reproducibility,Clancy:2019:OSIRRC,Frobe:2023:TireX,Ferro:2019:CENTRE}. In doing so, IIR-based shared tasks become more accessible to all practitioners in the IR community and result in less effort for organizers who would otherwise need to build much of this infrastructure themselves.

To demonstrate the utility of the resource, we re-deployed the Sim4IA Shared Task~\cite{Schaer:2025:Sim4IA-ArXiv,Sim4IA:2025:GitHubSharedTask} and this transforms participation in the shared task from a coding challenge into a streamlined process with minimal overhead. Our goal is to support a hosted version of \systemname{} for as long as possible (i.e., a minimum of at least 5 years), while simultaneously providing a fully containerized local deployment option for researchers requiring data privacy or offline access.\footnotemark[\getrefnumber{fn:searchsim}]\textsuperscript{,}%
\footnote{\label{fn:api_wrapper}\url{https://github.com/simint-ai/simiir-studio}} Moreover, should the community wish to become active developers, we welcome direct contributions to the repository. 


\section{Related Work}

Several frameworks have been developed to model user interactions with different levels of control. The SimIIR framework~\cite{Maxwell:2016:SIGIR-SimIIR,Zerhoudi:2022:CIKM-SimIIR2,Azzopardi:2024:SIGIR-AP} provides a modular architecture for simulating query generation, clicking, and stopping behaviors~\cite{Craswell:2008:WSDM-PositionBias,Dupret:2008:SIGIR-UBM,Maxwell:2015:CIKM-Stopping,Paakkonen:2017:IRJ-ValidatingSim}. More recently, the UserSimCRS~\cite{Bernard:2025:UserSimCRS,Afzali_2023} and UXSim~\cite{Zerhoudi:2025:CIKM} frameworks have integrated conversational and generative AI-based user models. While these tools are powerful, they are primarily code-centric libraries. They lack a unified graphical interface for component composition, requiring researchers to understand the system architecture prior to simulation. \systemname{} serves as an orchestration layer for different simulation frameworks. Currently,  SimIIR~3~\cite{Azzopardi:2024:SIGIR-AP} is the only supported framework due to the authors' familiarity with it and its modularity and established adoption.

In the broader IR ecosystem, platforms like TIREx~\cite{Frobe:2023:TireX}, repro\_eval \cite{Breuer:2021:repro_eval}, and other container-based evaluation infrastructures~\cite{Clancy:2019:OSIRRC} have standardized the execution of retrieval systems through Evaluation-as-a-Service (EaaS)~\cite{Hopfgartner:2018:evaluation}. \systemname{} differs by focusing on the design of the experimental subject itself---the user simulator---rather than the system under test. This distinction matters for interactive IR experiments where the simulator logic is a primary experimental variable. \systemname{} complements EaaS infrastructure, with the simulator bundle acting as a rerunnable companion artifact. More broadly, general-purpose experiment management tools (e.g., MLflow, Jupyter-based workflows) support pipeline tracking and reproducibility, but they do not provide a domain-specific interface for composing behavioral models from interacting simulation components, which is the core requirement for IIR simulation research.


\section{\systemname{} Design}
\systemname{} is designed to streamline the research cycle, from experiment design to analysis and finally to sharing. Its architecture is built to separate the user-facing interface from the underlying simulation engines, enabling flexibility and extensibility beyond the initial implementation using the SimIIR~3 framework. Throughout the remainder of this paper, we use the term \emph{simulation run} (or simply \emph{run}) to refer to a single execution of a simulation pipeline, which produces logs, behavioral traces, and evaluation outputs.

\subsection{System Features}
To facilitate ease of development as well as ease of use, \systemname{} is broken into several high-level features that we will describe before an in-depth architectural overview.

\subsubsection{Public Workbench}
\systemname{} is available via a public workbench at \url{https://iirsim.com}. The hosted instance supports the main workflows described in Section~\ref{sec:architecture}: interactive tutorial, playground runs, pipeline composition, experiment saving, and export of experiment bundles. This workbench is the main interaction point for shared tasks and less experienced researchers, as we expect more experienced researchers to focus on piloting pipelines before moving to large-scale runs with exported bundles. We plan to maintain the publicly hosted dashboard for as long as feasible and as interest dictates (e.g., by adding new shared tasks). Researchers requiring data privacy or institutional control can deploy \systemname{} on their own infrastructure using the files provided in the repository. 

\subsubsection{Open Repository and API Wrapper}
The \systemname{} API wrapper is publicly available at \url{https://github.com/simint-ai/simiir-studio}. While the public workbench is offered as a hosted service, the API wrapper provides programmatic access to all core functionality and can be integrated with custom interfaces. The wrapper functions as a communication layer between the underlying simulation framework (i.e., SimIIR~3) and client applications, supporting experiment bundle submission, batch execution, output collection, and result retrieval. The API wrapper allows researchers to build their own UIs or scripts that leverage the backend infrastructure. 

The API wrapper is the primary component that needs to be updated should additional simulation frameworks be supported. This was an intentional design decision to ensure that supporting new frameworks is straightforward and relatively painless. While this does mean that there is perhaps less specific support for SimIIR~3 (or other frameworks) in \systemname{} (i.e., in terms of UI elements), we believe that ensuring more general support will be beneficial in the long term for sustained impact and usage.

\begin{figure*}[t]
    \centering
    \includegraphics[width=\linewidth]{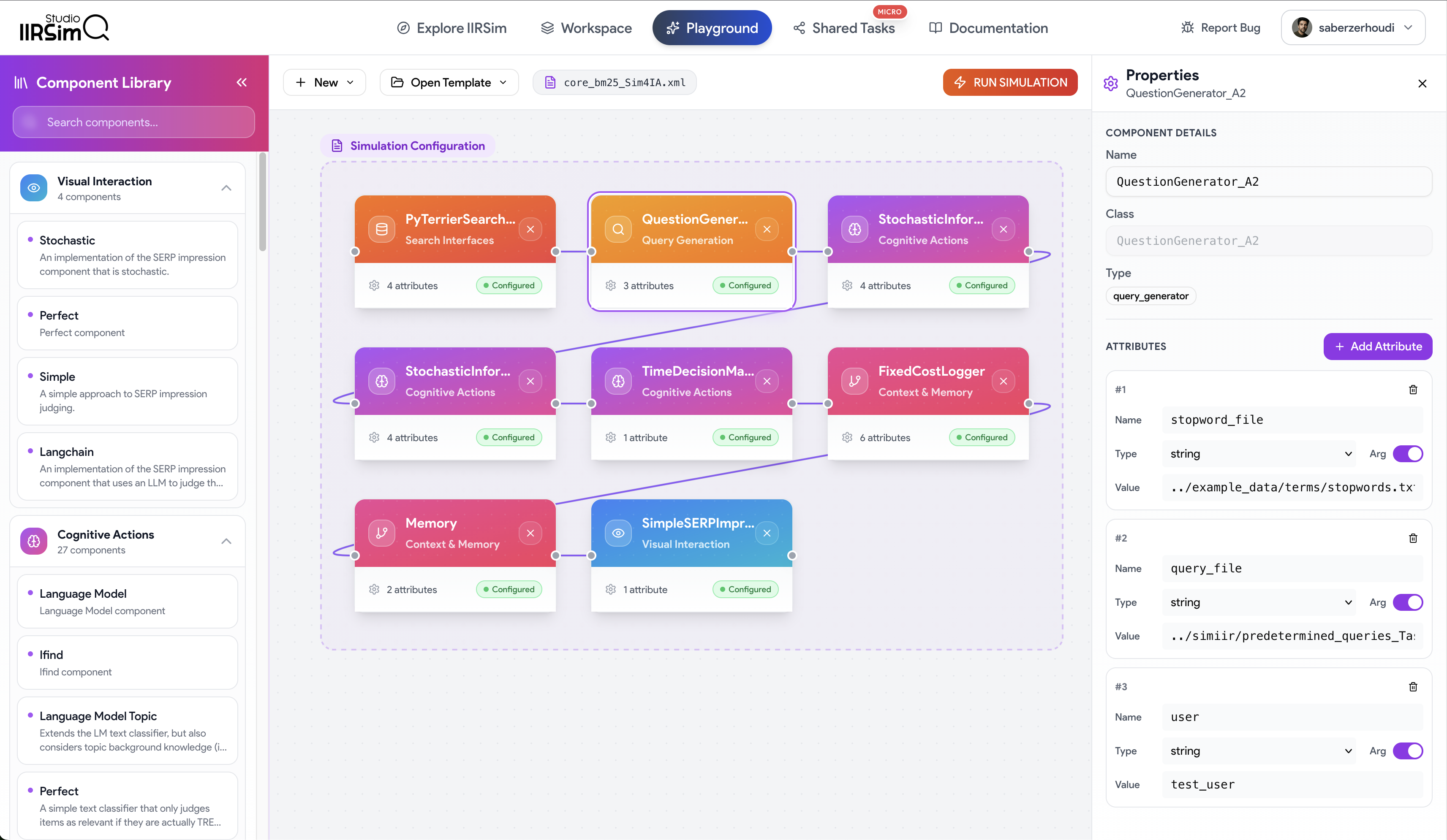}
    \caption{The visual pipeline composer. Each node represents a simulation component (e.g., query generator, stopping strategy) with its own configuration panel. Users connect components to define the simulator pipeline that will be executed.}
    \label{fig:core}
\end{figure*}

\subsubsection{Local Deployment}
The repository includes instructions for running \systemname{} locally through Docker, including a Docker Compose setup. Local deployment supports institutions requiring strict control over data, ensures the resource can outlive the hosted version, and allows experiments to be reproduced independently. Local deployment is a core supported mode with the same experiment bundle format and component loading contract.


\subsubsection{Reproducible Experiments}
\systemname{} uses an \emph{experiment bundle} as the unit of sharing and archiving. The bundle is intended to act as both an archival artifact as well as functional data for running experiments locally. The bundle format, described later, contains the necessary configuration details for replication.

\subsubsection{Environment Templates}
\systemname{} introduces the notion of an \emph{environment template}. An environment template bundles the dataset configuration, the search backend configuration, and baseline simulator pipelines into a named package that can be loaded in the UI. These templates can function as pre-packaged tutorials for helping new and returning users understand IIR simulation. 
Moreover, they provide a convenient method for organizers of shared tasks to ensure consistent environments for development and evaluation that can also be archived.

Environment templates and experiment bundles work together to support reproducibility: a template defines the fixed evaluation context (dataset, search backend, container versions), and the experiment bundle captures a specific simulation configuration within that context. When a user exports an experiment bundle, it includes a reference to the template version that was active, ensuring that the bundle can be re-executed against the same environment.

\begin{figure}[t]
  \centering
  \includegraphics[width=\linewidth]{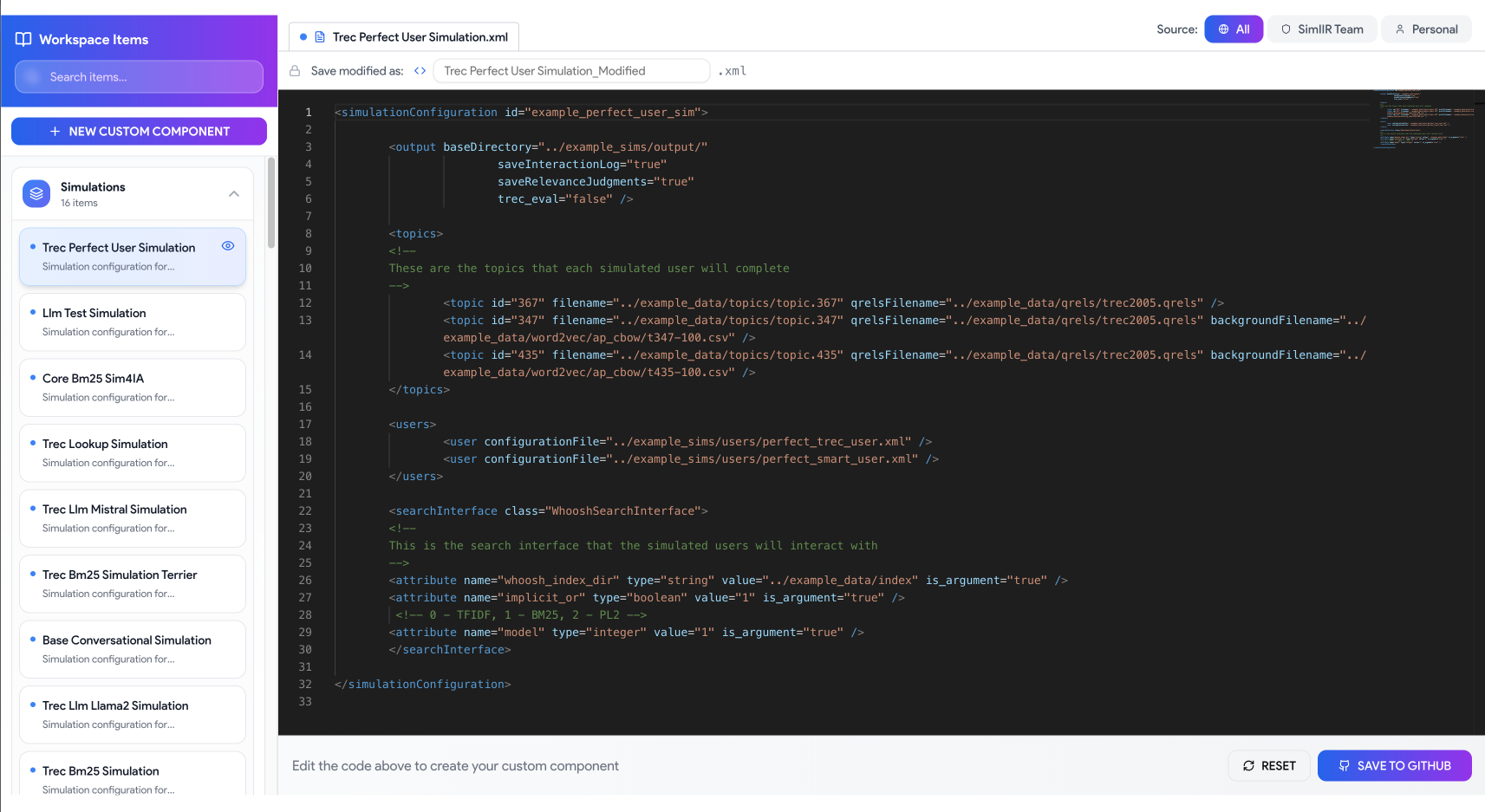}
  \caption{Custom component authoring and saving: in-browser code editor, category selection, and saved components appearing in the component library.}
  \label{fig:component_save}
\end{figure}

\begin{figure}[t]
  \centering
  \includegraphics[width=\linewidth]{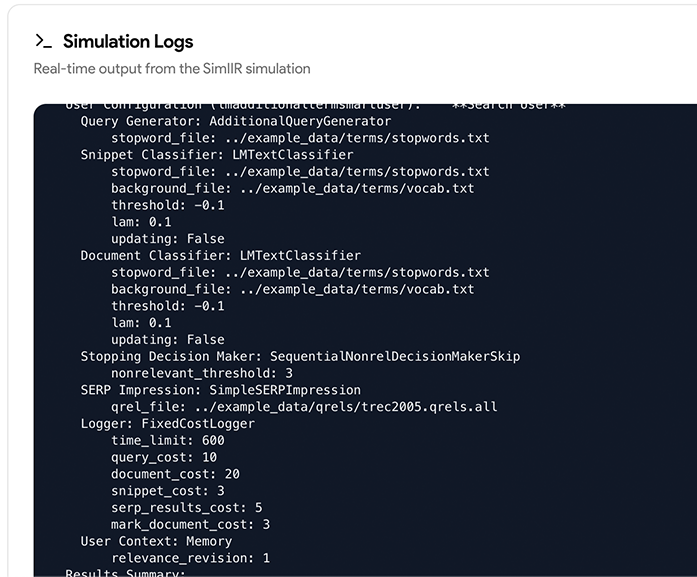}
  \caption{The playground results interface displaying simulation logs and behavioral traces. Users inspect these outputs to validate whether a configuration behaves as expected before exporting for large-scale runs.}
  \label{fig:run_experiment}
\end{figure}

\subsection{System Architecture}
\label{sec:architecture}

As shown in Figure~\ref{fig:architecture}, the system consists of three main layers: a frontend workbench for visual pipeline composition, an orchestration backend for experiment management and serialization, and an isolated execution layer based on Docker for reproducible runs.  

\subsubsection{Frontend Workbench}
\label{sec:frentend-workbench}
The frontend interface is designed to allow users to get started with simulation as quickly as possible without requiring heavy knowledge of simulation frameworks, while allowing more expressiveness as their confidence grows. There are many possible interfaces and there are plans to refine the interface based on additional user feedback, but the initial design is based on the existing UX Sim tool~\cite{Zerhoudi:2025:CIKM} for simplicity.

The \emph{pipeline composer} (Figure~\ref{fig:core}) provides a visual canvas where a simulator is built by connecting components. Each component has a configuration panel. The composer is how the user ties elements of the Complex Searcher Model~\cite{Maxwell:2018:ECIR} (or other behavioural models~\cite{Baskaya:2013:CIKM}) to the various elements of the simulation. For example, which query generator is used with a given search interface, which component interprets the search engine results, and which stopping component decides when the search is over. 

The \emph{playground} is intended to run short simulations for piloting and prototyping of larger scale simulation. To do so, it displays logs and simulated behavioural traces to conduct initial analyses (Figure~\ref{fig:run_experiment}). This allows the user to validate whether a given simulation configuration is useful or interesting before running large-scale experiments. In addition, the logged information allows users to debug any custom components that they designed. 

The \emph{interactive tutorial} is essentially a hybrid of the pipeline composer and playground that is intended to provide new users (whether to the tool or to simulation) introductory concepts with small runnable examples. The current tutorial is designed to help build understanding of the SimIIR~3 framework and how  changes alter simulation outcomes, but additional tutorials can be added easily as new frameworks are supported. 

The frontend also includes a \emph{component editor} (Figure~\ref{fig:component_save}) that allows users to author custom simulation components directly in the browser. This editor supports creating new Python classes that implement simulation component interfaces, with basic validation through linting and import checks. The component authoring workflow is described in detail in Section~\ref{sec:custom_components}.

\subsubsection{Orchestration Backend}

To ensure that simulation experiments are reproducible and archivable, the backend of \systemname{} maps UI actions to artifacts and experiment runs that can be archived and distributed. Moreover, it is responsible for ensuring that experiments run in the appropriate containers and acts as a first-level repository for experiments and environment templates. The backend can be accessed programmatically through the API wrapper, which enables large-scale experiments to be run after they have been prototyped. This is particularly useful for collaboration and distribution as one researcher can develop the simulation and another is able to run the experiments on private infrastructure to prevent data leakage. This functionality is supported through a process of \emph{experiment serialization}, which occurs when a pipeline is saved in a structured format that includes all relevant information for replication (e.g., component identifiers, random seeds and other parameters, versioned code references). The exported format is an \emph{experiment bundle} that we describe below.

\subsubsection{Experiment Bundle Format}
The experiment bundle is designed to be human-readable to facilitate auditing and analysis as well as to provide sufficient details to allow others to replicate the simulation through the API wrapper. The bundle describes the simulator pipeline as a directed graph of components, with explicit parameter settings and the version identifiers of the components. This includes the simulation framework version used for the run, the container image identifier, and the commit identifiers for any custom components. The bundle records the random seeds used for the run and identifies the environment template that provided the dataset and backend settings. Auxiliary files (e.g., session data and evaluation results) are stored separately and referenced within the configuration. Simulation outputs (logs, query files) can be exported through the API for analysis and storage.

\subsubsection{Isolated Execution with Docker}
To simplify development and maintenance, \systemname{} uses the same execution approach in hosted and local modes. This is accomplished by executing simulations through the use of coordinated Docker containers that include the simulation engine and the available search backends. Containerization supports repeatable environments across machines and simplifies aspects of dependency management for end users. While containerized setups can suffer from degradation over time, we mitigate this by ensuring that requirements specific to this project are self-contained in the repository rather than relying on external hosting of source files or data, which can limit utility when those components are no longer available. To further support long-term availability, we maintain official container images in a public registry that are tagged to specific versions of \systemname{}.

\paragraph{LLM Integration.} Given the increasing use of Large Language Models (LLMs) in simulation~\cite{Azzopardi:2024:SIGIR-AP,Bernard:2025:UserSimCRS}, \systemname{} supports remote LLM APIs and local LLM deployments. For remote APIs (e.g., OpenAI, Anthropic), users configure API keys through the component parameters, and the simulation container makes the outbound requests. For local LLMs, the Docker Compose configuration can be extended to include containers running local inference servers (e.g., Ollama, vLLM) that are accessed using the internal Docker network. This allows researchers to choose between the convenience of remote APIs and the reproducibility of local models.

\subsubsection{Reproducibility Contract}
\systemname{} supports reuse and replication with a clear and specific contract. If the same bundle is run with the same container image and the same component commit identifiers, then the experiment uses the same code versions and the same configuration. To promote deterministic outcomes, stochastic components require fixed seeds to be supported and must respect that seed. However, for experiments that rely on external services, such as remote LLM APIs, we cannot provide similar guarantees of repeatability. In such cases, the bundle still records which component version was used and which parameters were set, but the external service can return different outputs. This style of contract avoids absolute statements about experiment outcomes, rather, it makes the scope of replication explicit. Artifacts and settings can be preserved, as best as possible, for users who require strict control, while also allowing local execution.

\subsubsection{Version Policy for Templates}
As shared tasks can evolve over time (e.g., different TREC iterations), changing environment templates should not invalidate or break existing archived experiment bundles. To support this, each template is associated with specific version identifiers and points to specific container image identifiers. Accordingly, updating a template results in new version identifiers being generated, with the old versions remaining available. As exported experiment bundles reference the template version used for the experiment, replication is preserved.

\subsection{Supporting Custom Components}
\label{sec:custom_components}
Authoring custom simulation components (e.g., stopping decisions, query generators) is an integral to ensuring simulations reflect particular demographics. However, the process of editing and creating components in a low-level simulation framework (i.e., SimIIR~3) is not inherently the most accessible. To help users, we offer a visual editor in the provided UI (Figure~\ref{fig:component_save}) that allows them to author their own custom components. During creation, the user selects the category of component (currently those offered by SimIIR~3) to ensure that it can be appropriately integrated into the simulation framework and that it can be discovered in the simulation loader. Additionally, the editor allows for basic validation through standard Python linting and import checks.

\subsubsection*{GitHub-backed Saving and Branching}
\systemname{} uses GitHub to store and version custom components to more easily facilitate replication and distribution of code. Users log in to \systemname{} using GitHub OAuth, which means we do not store user passwords or other sensitive information. Upon authentication, the system prepares a fork of the simulation framework (e.g., SimIIR~3) that acts as a personal component store, which is cached by the backend. When the user saves a component, the backend writes the file into the correct directory inside the repository, creates a commit, and pushes that commit to the user's fork. By integrating with Github, users are provided with a normal git history that can be audited and verified as well as providing a specific commit for the experiment bundle to use. This allows replication since there are explicitly managed versions (using commit hashes) rather than relying on specific naming conventions or other styles of versioning. Moreover, users can then merge useful simulation components back to the source repository as a means of contributing to the community.


\subsubsection*{Dependency Handling}
Custom components may require extra libraries, and supporting them is non-trivial, especially while maintaining security and privacy. The hosted \systemname{} provides a base environment with common IR and ML libraries (e.g., \emph{numpy}, \emph{scikit-learn}, \emph{openai}). Hosted runs are designed to remain stable for many users, so arbitrary system-level dependency installation is not supported. Additional dependencies will be considered should there be wide appeal for them.
In a local deployment, users can modify the container build files to include additional Python packages or local LLMs using provided instructions.

\subsubsection*{Parameter Exposure in the UI}
A low-code interface requires a stable mapping from code parameters to UI fields. \systemname{} exposes parameters through a simple contract: if the component constructor uses typed arguments and default values, the UI can infer a basic form. The system also supports an optional schema file stored with the code, which defines field names, descriptions, and permitted values. This keeps the UI readable and reduces accidental misuse.

\subsubsection*{Runtime Injection via \texttt{PYTHONPATH}}
A saved custom component becomes usable without rebuilding the SimIIR~3 engine. The execution container mounts the user's component repository into the container filesystem and adds the path to \texttt{PYTHONPATH}. 
SimIIR~3 uses dynamic import to discover component classes available on \texttt{PYTHONPATH}, which it uses to instantiate the class referenced in the configuration.
This approach supports short edit-run cycles to foster better understanding between cause and effect when making simulation changes (i.e., the user can make changes to a component and rerun an experiment without requiring a restart).

\section{Intended Use Cases}
In this section, we highlight how \systemname{} can help three different audiences: those new to simulation or with less technical skills, experienced researchers, and those conducting collaborative events (e.g., tutorials, workshops, shared tasks). 

\subsection{New to Simulation}
The interactive tutorial hosted on the public workbench introduces core simulation blocks through small runnable examples. The playground allows users to load a template pipeline and change parameters to observe how behavior changes, building understanding of how components interact without requiring deep technical knowledge of the underlying framework. As users develop familiarity with simulation primitives, the component editor allows them to author custom components, include them in a pipeline, and validate them through the playground. This progression is designed to lower the initial hurdle for less technical users and to support the gradual adoption of simulation as a research methodology.

\subsection{Experienced Researchers}
The pipeline composer and experiment playground support experimental design and validation, allowing experts to refine a simulation configuration before scaling up. After validation, an experiment bundle can be exported and executed at scale through the API wrapper, which collects experiment outputs programmatically. This separation of prototyping and execution is particularly useful for collaboration, as one researcher can design the simulation on the hosted workbench, while another runs the experiments on private infrastructure with additional dependencies as needed. 

\subsection{Shared Task and Tutorial Organizers}
Organizers define a template environment that pins the dataset, the search backend configuration, and baseline simulator pipelines. Participants copy the template and modify only the simulator components, with submissions represented as experiment bundles that the organizer can rerun using the same template version.

To demonstrate this workflow, we re-deployed one of the Sim4IA micro shared-tasks~\cite{Schaer:2025:Sim4IA-ArXiv,Sim4IA:2025:GitHubSharedTask} within \systemname{}. Concretely, this involved creating a template environment that pinned the Sim4IA dataset and search backend, defining a baseline pipeline, and representing the task's evaluation criteria as a rerunnable configuration. The re-deployment was conducted by the authors to validate feasibility rather than by external participants, but it confirmed that the template and bundle mechanisms can support the shared-task workflow end-to-end. Additionally, \systemname{} serves as the backbone for a CHIIR 2026 tutorial~\cite{zerhoudi2026simulation}. Both deployments guided the design of the template and bundle features presented herein.

\section{Future Directions}
The \systemname{} public workbench is the intended main access point for most users of \systemname{}, at least initially, and so we intend for it to be available as long as possible with a current horizon of at least five years. To facilitate community involvement and local deployment, we license \systemname{} under the permissive open-source MIT license and provide documentation for users to host their own version workbench and the underlying infrastructure. To help ensure replication of experiments, the project documentation provides a relevant citation and versioned release tags to support stable references in papers and shared tasks.

While \systemname{} currently supports only the SimIIR~3 simulation framework, this was a choice made based on familiarity and as a starting point. The design of \systemname{} is such that if the community wishes to have other frameworks supported then it can be facilitated through collaborative endeavours and support from the paper authors. Indeed, \systemname{} is intended to be as framework agnostic as possible to ensure that subsequent modifications do not break replicability.

We plan to further improve the user experience through feedback from use of \systemname{} in tutorials and shared tasks. Moreover,  we plan to conduct user studies to better understand how researchers of different experience levels can be supported. 

\section{Limitations and Considerations}



\subsubsection*{Security and Privacy.} Running user-provided code raises safety and privacy questions. The hosted instance executes simulation runs in isolated containers and is intended for research code. Authentication is handled through GitHub OAuth, meaning \systemname{} does not store user credentials directly. However, sensitive data scenarios are better served through local deployment, where institutions can enforce their own security and network policies.

When users configure external API keys for LLM access, they should assume that prompts or document content may be sent to that provider during simulation execution. For experiments involving proprietary or sensitive data, local deployment with locally-hosted LLMs is recommended. The containerized architecture supports this configuration through Docker Compose extensions.

\subsubsection*{Platform Scope.} \systemname{} currently focuses on SimIIR~3 as the execution core. While the architecture separates the workbench from the execution layer to allow additional simulation engines to be integrated, the current release only guarantees the SimIIR-based pathway. Community contributions to support additional frameworks are welcome.

\subsubsection*{Adoption Requirements.} While \systemname{} was built to maximize the ability for simulations to be reproduced and replicated, the design decisions were based on the expertise and familiarity of the authors. Accordingly, the most value is gained by those with familiarity of the underlying tooling (e.g., Docker, Github). The result is that further work remains to ensure that \systemname{} and the results it produces are replicable by as much of the community as possible and not just those with similar background to the authors.

\section{Conclusion}
\systemname{} is a released resource that addresses the infrastructure gap between simulation engines and reproducible simulation experiments in IR. Built on top of SimIIR~3, it connects visual pipeline design, Git-backed component versioning, containerized execution, and versioned experiment bundles into a single workflow that supports both novice and expert researchers. As user simulation becomes increasingly central to IR evaluation, particularly with the growth of LLM-based user models and conversational search, the need for shared infrastructure that supports experiment design, execution, and replication will continue to grow. \systemname{} provides a foundation for this infrastructure and is available for community use and contribution.

%
%
%
\bibliographystyle{ACM-Reference-Format}
\bibliography{refs}

\end{document}